# Absence of the impurity-induced magnetic order in the electron-doped high-$T_c$ cuprates $Pr_{0.86}LaCe_{0.14}Cu_{1-y}(Zn, Ni)_yO_4$

Risdiana[a*], T. Adachi[a], Y. Koike[a] and I. Watanabe[b], K. Nagamine[c]

[a]Department of Applied Physics, Tohoku University, 6-6-05, Aoba, Aramaki, Aoba-ku, Sendai 980-8579, Japan
[b]Advanced Meson Science Laboratory, RIKEN, Wako 351-0198, Japan
[c]Muon Science Laboratory, Institute of Materials Structure Science, KEK-IMSS, 1-1 Oho, Tsukuba 305-0801, Japan

**Abstract**

Zero-field muon-spin-relaxation measurements have been carried out in order to investigate the Zn- and Ni-substitution effects on the Cu-spin dynamics in the electron-doped $Pr_{0.86}LaCe_{0.14}Cu_{1-y}(Zn, Ni)_yO_{4+\alpha-\delta}$ with y = 0, 0.01, 0.02, 0.05 and different values of the reduced oxygen content $\delta$ ($\delta \leqq 0.09$). For the samples with y = 0 and very small $\delta$ values of $\delta < 0.01$, a muon-spin precession due to the formation of a long-range antiferromagnetic order has been observed at low temperatures below ~ 5 K. For the moderately oxygen-reduced samples of $0.01 \leqq \delta \leqq 0.09$, on the contrary, no muon-spin precession has been observed and the temperature dependence of the spectra is similar to one another regardless of the y value. That is, no impurity-induced slowing down of the Cu-spin fluctuations has been detected, which is very different from the results of the hole-doped high-$T_c$ cuprates. The reason is discussed.

## 1. Introduction

Since the discovery of the superconductivity in the electron-doped high-$T_c$ cuprate $Nd_{2-x}Ce_xCuO_4$ [1], the electron-hole doping symmetry on the high-$T_c$ superconductivity has attracted great interest in relation to the mechanism of the superconductivity. Some properties in the electron-doped superconductors have been found to be different from those in the hole-doped superconductors. In the hole-doped system, for instance, holes are introduced into the O-2p state, giving rise to the magnetic frustration in the Cu-spin ordering [2]. In the electron-doped system, on the other hand, electrons are introduced into the Cu-3d state, diluting Cu spins without giving rise to any magnetic frustration [3]. To give another example, the superconductivity in the electron-doped $Nd_{2-x}Ce_xCuO_4$ is suppressed through the Ni substitution for Cu more markedly than through the Zn substitution [4], which is contrary to the result in the hole-doped high-$T_c$ cuprates.

In the hole-doped $La_{2-x}Sr_xCu_{1-y}(Zn, Ni)_yO_4$, we have recently found from the muon-spin-relaxation ($\mu$SR) measurements that the partial substitution of Zn for Cu develops a magnetic order associated with the stripe order of spins and holes [5] more markedly than the Ni substitution [6,7]. This is interpreted as a result of the strong pinning effect of Zn on the dynamically fluctuating stripes of spins and holes [8,9]. Similar effects of Zn on the Cu-spin dynamics has been observed in the hole-doped $Bi_2Sr_2Ca_{1-x}Y_x(Cu_{1-y}Zn_y)_2O_{8+\delta}$ [10,11] and $YBa_2Cu_{3-2y}Zn_{2y}O_{7-\delta}$ [12] as well. Accordingly, we have investigated the Zn- and Ni-substitution effects on the Cu-spin dynamics in the electron-doped $Pr_{0.86}LaCe_{0.14}Cu_{1-y}(Zn, Ni)_yO_4$ with y = 0, 0.01, 0.02, 0.05 from the μSR measurements.

---

* Corresponding author. Tel.: +81-22-795-7977; Fax: +81-22-795-7975; e-mail: risdiana@teion.apph.tohoku.ac.jp

## 2. Experimental

Polycrystalline samples of $Pr_{0.86}LaCe_{0.14}Cu_{1-y}(Zn, Ni)_yO_{4+\alpha-\delta}$ with y = 0, 0.01, 0.02, 0.05 were prepared by the ordinary solid-state reaction method as follows [13,14]. Raw materials of dried $La_2O_3$, $Pr_6O_{11}$, $CeO_2$, CuO and ZnO or NiO powders were mixed in a stoichiometric ratio and prefired in air at 900°C for 20 h. The prefired materials were reground and pressed into pellets of 10 mm in diameter, and sintered in air at 1100°C for 16 h with repeated regrinding. As-grown samples of $Pr_{0.86}LaCe_{0.14}Cu_{1-y}(Zn, Ni)_yO_{4+\alpha}$ were post-annealed in flowing Ar gas of high purity (6N) at 950°C for 10 h in order to remove the excess oxygen at the apical site. The reduced oxygen content δ was estimated from the weight change before and after annealing. All of the samples were checked by the powder x-ray diffraction measurements to be single phase. Both electrical resistivity and magnetic susceptibility were measured to check the quality of the samples and to determine the superconducting transition temperature, $T_c$.

The μSR measurements were performed at the RIKEN-RAL Muon Facility at the Rutherford-Appleton Laboratory in the UK, using a pulsed positive surface muon beam at temperatures down to 2 K.

## 3. Results and discussion

Figure 1 shows the zero-field (ZF) μSR time spectra of $Pr_{0.86}LaCe_{0.14}CuO_{4+\alpha-\delta}$. As the superconductivity in the electron-doped system is affected by the reduced oxygen content δ, the spectra are grouped into 3 classes with different δ values; very small δ of δ<0.01, small δ of 0.01≦δ<0.04 and large δ of 0.04≦δ≦0.09. The samples of the small and large δ show superconductivity with $T_c$ ranging from 15 K to 17 K (the average $T_c$~16 K) and from 18 K to 22 K (the average $T_c$~20 K), respectively, while the samples of the very small δ are not superconducting above 4.2 K. For every sample, a Gaussian-like behavior is observed at high temperatures above ~ 100 K due to the randomly oriented nuclear spins, and an exponential-like depolarization of muon spins is observed at low temperatures below ~ 50 K. For the sample of the very small δ, a muon-spin precession due to the formation of a long-range antiferromagnetic order is observed at low temperatures below ~ 5 K. For the samples of the small and large δ, on the other hand, the spectra seem to be independent of temperature at low temperatures below 20 K and no muon-spin precession is observed, indicating the absence of a long-range magnetic order above 2 K. The temperature-dependent change of the spectra above 20 K is regarded as being due to static random magnetism of the $Pr^{3+}$ moments [15]. The $\delta$-independent behavior of the spectra in the superconducting samples of the small and large $\delta$ indicates that the $\delta$ value does not affect the Cu-spin dynamics so much except for the sample of the very small $\delta$.

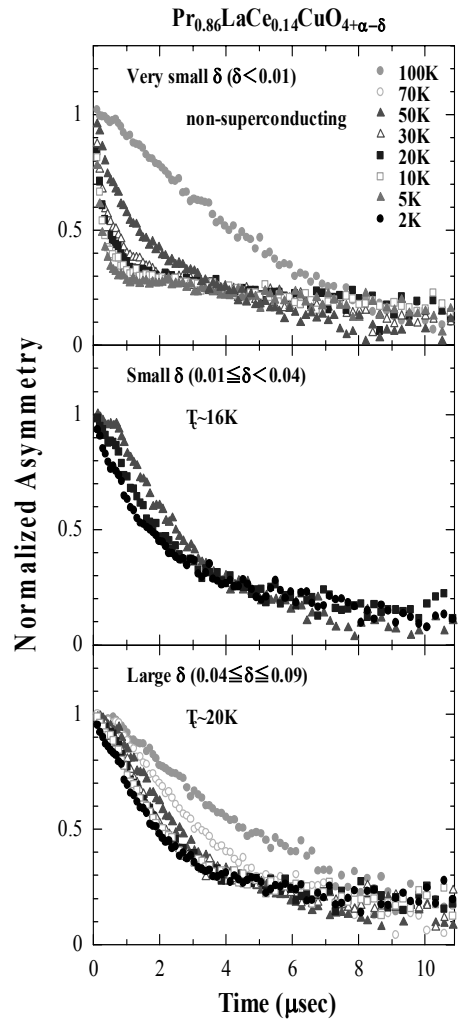

Figure 1 : ZF-μSR time spectra of $Pr_{0.86}LaCe_{0.14}CuO_{4+\alpha-\delta}$ of three groups with different δ values, very small δ (δ<0.01), small δ (0.01≦δ<0.04) and large δ (0.04≦δ≦0.09), at various temperatures down to 2 K. The $T_c$ value in the figure is the average one for each group.

Figure 2 shows the ZF-$\mu$SR time spectra of $Pr_{0.86}LaCe_{0.14}Cu_{1-y}(Zn, Ni)_yO_{4+\alpha-\delta}$ with various y and $\delta$ values. The temperature dependence of the spectra is similar to one another. No impurity-induced slowing down of the Cu-spin fluctuations is detected for moderately oxygen-reduced samples of the small and large $\delta$, which is very different from the results of the hole-doped high-$T_c$ cuprates [6-12]. This may be understood in two ways. First, there may be no dynamically fluctuating stripes of spins and electrons in the electron-doped system, because the dynamical stripes lead to the impurity-induced magnetic order in the hole-doped system [8,9]. Second, $Pr_{0.86}LaCe_{0.14}CuO_4$ may be situated in the overdoped regime, because Cu spins fluctuate so fast that the measured lowest temperature of 2 K is too high to detect the slowing down of the Cu-spin fluctuations in the overdoped regime of the hole-doped $La_{2-x}Sr_xCuO_4$ [16]. In any case, the Cu-spin dynamics within the μSR time window (typically from $10^{-6}$ to $10^{-11}$ sec) in the electron-doped system may not be so affected by any impurity as follows. Generally, impurities tend to make carriers localized. In the hole-doped system, as doped holes give rise to the magnetic frustration, whether holes are mobile or localized may affect the Cu-spin dynamics strongly, leading to large effects of impurities on the Cu-spin dynamics. In the electron-doped system, on the other hand, doped electrons give rise to no magnetic frustration and only dilute Cu spins. Therefore, whether doped electrons are mobile or localized may not affect the Cu-spin dynamics so much, leading to no significant effects of impurities on the Cu-spin dynamics.

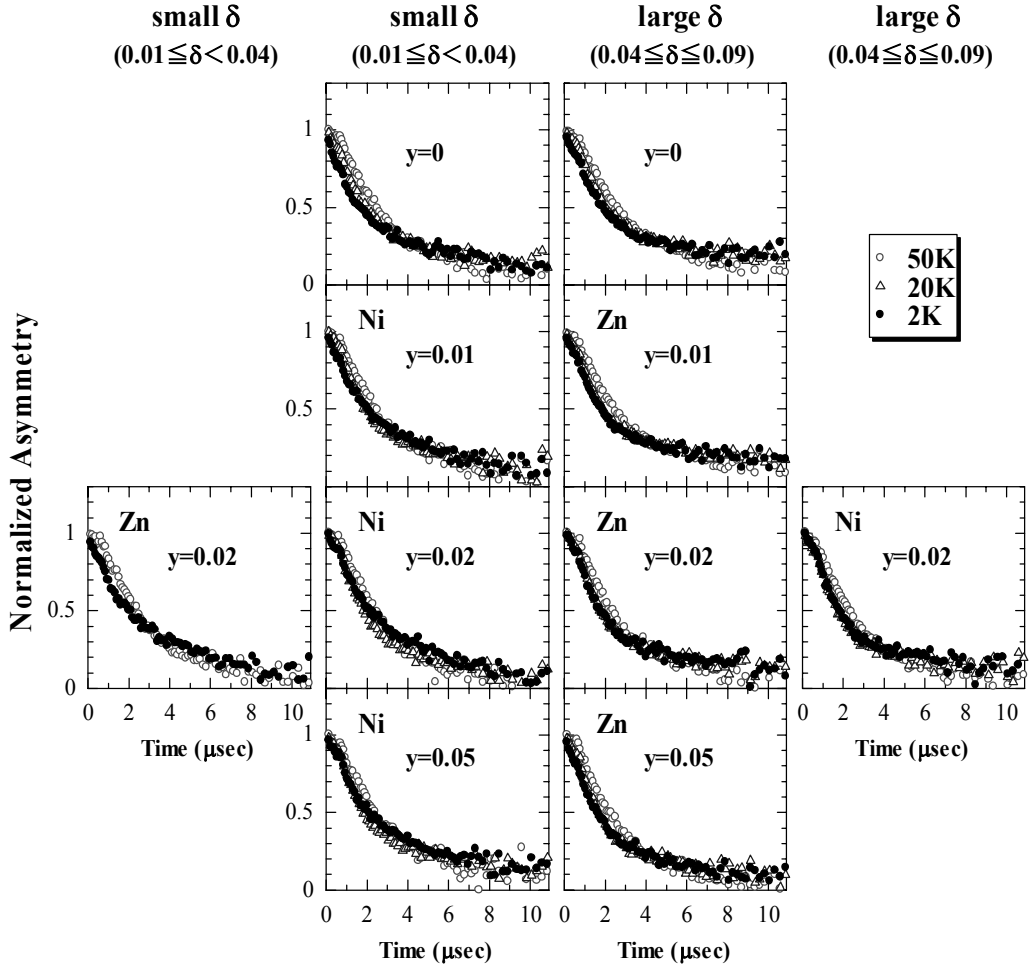

Figure 2 : ZF–$\mu$SR time spectra of $Pr_{0.86}LaCe_{0.14}Cu_{1-y}(Zn, Ni)_yO_{4+\alpha-\delta}$ with y = 0, 0.01, 0.02 and 0.05 at 50 K, 20 K and 2 K for small and large $\delta$.

## 4. Summary

We have investigated the Zn- and Ni-substitution effects on the Cu-spin dynamics from the ZF-μSR measurements in $Pr_{0.86}LaCe_{0.14}Cu_{1-y}(Zn, Ni)_yO_{4+\alpha-\delta}$ with y = 0, 0.01, 0.02, 0.05 and various δ values at temperatures down to 2 K. For the samples with y = 0 and very small δ values of δ < 0.01, a muon-spin precession due to the formation of a long-range antiferromagnetic order has been observed at low temperatures below ~ 5 K. On the contrary, no impurity-induced slowing down of the Cu-spin fluctuations has been detected for the moderately oxygen-reduced samples of $0.01 \leqq \delta \leqq 0.09$, which is very different from the results in the hole-doped high-$T_c$ cuprates. This may be understood in two ways. First, there may be no dynamically fluctuating stripes of spins and electrons. Second, $Pr_{0.86}LaCe_{0.14}CuO_4$ may be situated in the overdoped regime. In any case, the Cu-spin dynamics within the μSR time window in the electron-doped system may not be so affected by any impurity because of no magnetic frustration effect of doped electrons.

## Acknowledgments

The authors thank M. Kato and N. Oki for helpful discussions.